\documentclass[10pt,conference]{IEEEtran}
\IEEEoverridecommandlockouts
\pdfoutput=1
\usepackage{cite}
\usepackage{amsmath,amssymb,amsfonts}
\usepackage{algorithmic}
\usepackage{algorithm,algorithmic}
\usepackage{url}
\usepackage{hyperref}
\usepackage{tikz}
\usepackage{graphicx}
\usepackage{textcomp}
\usepackage{xcolor}
\usepackage{soul}
\usepackage{amssymb}
\usepackage{systeme,mathtools}

\usepackage[outdir=./]{epstopdf}

\usepackage[normalem]{ulem}

\usepackage{pgfplots}
\pgfplotsset{compat=newest} 
\usepgfplotslibrary{units} 
\pgfplotsset{width=9cm,compat=1.18}
\usepgfplotslibrary{statistics}
 \usetikzlibrary{
        pgfplots.groupplots,
    }
    \pgfplotsset{
        my addplot style/.style={
            hist,
        },
    }
\usepackage{tikzpagenodes}


 \usepackage{mathrsfs}
 \usetikzlibrary{arrows}
\usepgfplotslibrary{groupplots}
\def\BibTeX{{\rm B\kern-.05em{\sc i\kern-.025em b}\kern-.08em
    T\kern-.1667em\lower.7ex\hbox{E}\kern-.125emX}}

\definecolor{orcidlogocol}{HTML}{A6CE39}
\tikzset{
  orcidlogo/.pic={
    \fill[orcidlogocol] svg{M256,128c0,70.7-57.3,128-128,128C57.3,256,0,198.7,0,128C0,57.3,57.3,0,128,0C198.7,0,256,57.3,256,128z};
    \fill[white] svg{M86.3,186.2H70.9V79.1h15.4v48.4V186.2z}
                 svg{M108.9,79.1h41.6c39.6,0,57,28.3,57,53.6c0,27.5-21.5,53.6-56.8,53.6h-41.8V79.1z M124.3,172.4h24.5c34.9,0,42.9-26.5,42.9-39.7c0-21.5-13.7-39.7-43.7-39.7h-23.7V172.4z}
                 svg{M88.7,56.8c0,5.5-4.5,10.1-10.1,10.1c-5.6,0-10.1-4.6-10.1-10.1c0-5.6,4.5-10.1,10.1-10.1C84.2,46.7,88.7,51.3,88.7,56.8z};
  }
}

\newcommand\orcid[1]{\href{https://orcid.org/#1}{\mbox{\scalerel*{
\begin{tikzpicture}[yscale=-1,transform shape]
\pic{orcidlogo};
\end{tikzpicture}
}{|}}}}

\newcommand\myineqlb{\stackrel{\mathclap{\normalfont\mbox{(b)}}}{\leq}}
\newcommand\myineqla{\stackrel{\mathclap{\normalfont\mbox{(a)}}}{\leq}}

\definecolor{clr1}{rgb}{0.0, 0.0, 1.0}
\definecolor{clr2}{rgb}{0.96, 0.73, 1.0}
\definecolor{clr3}{rgb}{1.0, 0.01, 0.24}
\definecolor{clr4}{rgb}{0.0, 0.5, 0.0}
\definecolor{clr5}{rgb}{1.0, 0.49, 0.0}
\definecolor{clr6}{rgb}{0.1, 0.1, 0.44}
\author{\IEEEauthorblockN{Vida Ranjbar\IEEEauthorrefmark{1}, Robbert Beerten\IEEEauthorrefmark{1}, Marc Moonen\IEEEauthorrefmark{1}, Sofie Pollin\IEEEauthorrefmark{2}}
\IEEEauthorblockA{\IEEEauthorrefmark{1}\textit{Department of Electrical Engineering, KU Leuven, Belgium} \\
\IEEEauthorrefmark{2}\textit{IMEC, Kapeldreef 75, 3001 Leuven, Belgium} \\
Corresponding Author: \{vida.ranjbar\}@kuleuven.be}}

\begin{document}

\title{Sequential Processing in Cell-free Massive MIMO Uplink with Limited Memory Access Points
}

\maketitle%
\begin{abstract}
Cell-free massive multiple-input multiple-output (MIMO) is an emerging technology that will reshape the architecture of next-generation networks. 
This paper considers the sequential fronthaul, whereby the access points (APs) are connected in a daisy chain topology with multiple sequential processing stages.
With this sequential processing in the uplink, each AP refines users' signal estimates received from the previous AP based on its own local received signal vector. While this processing architecture has been shown to achieve the same performance as centralized processing, the impact of the limited memory capacity at the APs on the store and forward processing architecture is yet to be analyzed. Thus, we model the received signal vector compression using rate-distortion theory to demonstrate the effect of limited memory capacity on the optimal number of APs in the daisy chain fronthaul. Without this memory constraint, more geographically distributed antennas alleviate the adverse effect of large-scale fading on the signal-to-interference-plus-noise-ratio (SINR). 
 However, we show that in case of limited memory capacity at each AP, the memory capacity to store the received signal vectors at the final AP of this fronthaul becomes a limiting factor. In other words, we show that when deciding on the number of APs to distribute the antennas, there is an inherent trade-off between more macro-diversity and compression noise power on the stored signal vectors at the APs. Hence, the available memory capacity at the APs significantly influences the optimal number of APs in the fronthaul.

\end{abstract}
\section{Introduction}
Massive multiple-input multiple-output (MIMO) is one of the critical enablers for next-generation mobile networks. It promises spectral efficiency (SE), energy efficiency (EE), and reliability and allows for low-cost hardware at both receiver and
transmitter \cite{ErikMassive_MIMO_NG}. 
Massive MIMO provides additional degrees of freedom in the 
spatial domain, allowing it to
separate users spatially rather than via time and frequency scheduling.
This reuse of time and frequency resources dramatically increases the average throughput.
Cell-free massive MIMO has attracted a lot of academic attention recently as a promising massive MIMO technology for the future generation of wireless networks due to its ability to mitigate the adverse effect of large-scale fading on the users' signal-to-interference-plus-noise-ratio (SINR) and to provide uniform service to all users \cite{Ngocellfree_vs_smallcells}. In such a network, the antennas are distributed among the access points (APs), and all or a few nearby APs will serve each user. As the serving APs are selected based on their vicinity to the user, the cell-edge phenomenon and poor coverage of traditional cellular networks disappear. 


In cell-free massive MIMO networks, the APs cooperate to serve the users with direct information exchange, e.g., in a sequential fronthaul topology\cite{shaik2020,Shaik2021,ke_Dmimo_kalman,ranjbar2022} or indirectly through a central processing unit (CPU) in a star fronthaul topology \cite{Ngocellfree_vs_smallcells,bjornsonscalable,makingbjornson}. In the uplink of a cell-free massive MIMO network with a daisy chain fronthaul topology, each AP estimates the users' signal and sends the local estimates to the next AP in the sequence. In this way, the users' signal estimates are refined through the daisy chain fronthaul. Hence, sequential processing in a daisy chain fronthaul topology requires the AP to store their received signal vector in the memory until they receive the corresponding information from the previous AP in the sequence. In \cite{Shaik2021}, it is proven that with sequential processing at each AP connected in a daisy chain fronthaul topology, and for the same number of exchanged scalars on each chunk of fronthaul, the minimum mean square error (MMSE) optimal solution 
can be achieved in the last AP. However, the memory constraint at each AP in the sequential fronthaul is 
neglected. 

Non-idealities such as limited bandwidth fronthaul links, hardware impairment, and low-resolution analog-to-digital converters (ADCs) in both cellular and cell-free massive MIMO networks are discussed in\cite{HardwareBjörnson,HuLRADC,optimalADC_verenzuela,Xiongletter,YouzhiBitallocation,Bashar_2021_uniform_q,bashar_EE_uq_2019,masoumiperformance}, among others.
In \cite{HardwareBjörnson}, it is shown under which scenarios the correlation between the distortion vector elements in a massive MIMO network with hardware impairment has a negligible impact on the users' SE. The ADC bit allocation among antennas in a cell-free massive MIMO network is discussed in \cite{optimalADC_verenzuela,HuLRADC,Xiongletter,YouzhiBitallocation}. In \cite{optimalADC_verenzuela}, the SE and EE maximization problems are formulated as a function of the number of ADC bits used to represent the antenna signals, subject to a constraint on the total number of bits or power consumption. In \cite{YouzhiBitallocation}, adaptive intra-AP and inter-AP bit allocation for ADCs is considered. In \cite{Bashar_2021_uniform_q,bashar_EE_uq_2019}, the impact of limited capacity fronthaul links on the users' SE and EE in cell-free massive MIMO uplink is investigated.


The number of bits to quantize the received signal vector is a vital cost metric for the ADC, in 
the case of limited fronthaul capacity and when the vector needs to be stored in memory in a network with sequential fronthaul topology, such as in this paper. The sequential fronthaul topology is studied in, e.g. \cite{shaik2020,Shaik2021,ke_Dmimo_kalman,ranjbar2022}. However, these works do not consider the limited capacity of the memory at the APs.
This paper studies sequential processing for uplink users' signal estimation in a cell-free massive MIMO network with a daisy chain fronthaul topology under the realistic assumption of a limited memory capacity constraint at each AP.

\textbf{Contribution}: To the best of the authors' knowledge, this paper is the first to address the problem of limited memory capacity availability at the APs due to the sequential processing in the cell-free massive MIMO with daisy chain fronthaul. First, we use a tractable model based on rate-distortion theory to model the compression of the received signal vectors in the memory of the APs. Second, using two limited memory capacity models, the effect of limited memory capacity on the optimal number of APs in the daisy chain fronthaul topology is quantified in the simulation section.  
\subsection{Notation}
We denote vectors and matrices with boldface lower-case and upper-case letters, respectively. Transpose and conjugate transpose operations are denoted by superscripts $^{\text{T}}$ and $^{\text{H}}$, respectively. A circularly symmetric complex Gaussian distribution with covariance matrix $\mathbf{X}$ is represented as $\mathcal{C}\mathcal{N}(0, \mathbf{X})$. Symbol $\mathbb{E}\{\mathbf{x}\}$ denotes the mean of $\mathbf{x}$. $\mathcal{H}(\mathbf{x})$ is the differential entropy of $\mathbf{x}$, and $I(\mathbf{x};\hat{\mathbf{x}})$ is the mutual information between $\mathbf{x}$ and $\hat{\mathbf{x}}$. The Euclidean norm of $\mathbf{x}$ is shown as $\lVert \mathbf{x} \rVert$. We use $\text{diag}(\mathbf{X})$ to signify the elements on the main diagonal of $\mathbf{X}$ and $\text{diag}(\mathbf{x})$ for a diagonal matrix with $\mathbf{x}$ as its main diagonal. Furthermore, $\mathbf{X}=\text{blkdiag}(\mathbf{X}_1,\hdots,\mathbf{X}_L)$ is a block-diagonal matrix with matrices  $\mathbf{X}_i~i =\{ 1,\dots,L\}$
as diagonal blocks. $\mathbf{X}^{1/2}$ is the square-root of $\mathbf{X}$. For two matrices $\mathbf{A}$ and $\mathbf{B}$, $\mathbf{A}\succeq\mathbf{B}$ means that $\mathbf{A}-\mathbf{B}$ is positive semi-definite. Finally, tr(${\mathbf{X}}$) denotes the trace of matrix ${\mathbf{X}}$.


\section{System model and Problem statement}\label{sec2}
Distributed processing in cell-free massive MIMO networks is a necessity as it avoids overloading a single AP with massive computations, and it
enables truly scalable implementations and, hence, large-scale deployments.
 Distributed processing has become even more attractive because of the growing interest in the sequential fronthaul topology\cite{Shaik2021,shaik2020,ke_Dmimo_kalman,ranjbar2022}. In this paper, we consider distributed uplink signal estimation using the least-squares (LS) method in a cell-free massive MIMO network with limited memory APs. There are $L$ APs, each having $N$ antennas, connected in a daisy chain fronthaul topology, serving $K$ single antenna users.
\subsection{Recursive least-squares (RLS) for uplink signal estimation}\label{secII-A}
The received signal vector at AP $l$ in the uplink is given as follows:
\begin{equation}
\mathbf{y}_l=\mathbf{H}_l\mathbf{s}+\mathbf{n}_l,
    \label{eq1}
\end{equation}
where $\mathbf{s}\sim \mathcal{C}\mathcal{N}(0,p \mathbf{I}_K)$ is the users' signal, $\mathbf{H}_l\in \mathbb{C}^{N\times K}$ and $\mathbf{n}_l\sim \mathcal{C}\mathcal{N}(\mathbf{0},\sigma^2\mathbf{I}_N)$ are the local channel matrix and noise vector at AP $l$, respectively. The channel vector between user $k$ and AP $l$ is drawn from a correlated Rayleigh distribution, i.e. $\mathbf{H}_{l[:,k]}\sim\mathcal{C}\mathcal{N}(\mathbf{0},\mathbf{R}_{kl})$, where subscript $[:,k]$ denotes the $k^{th}$ column of $\mathbf{H}_l$ and $\beta_{kl}=\text{tr}(\mathbf{R}_{kl})/{N}$. We assume a block fading model in which the channel matrix $\mathbf{H}_l$  remains constant in a coherence interval of $\tau_c=B_cT_c$ samples, with $T_c$ and $B_c$ the coherence time and coherence bandwidth of the channel, respectively \cite{marzetta_larsson_yang_ngo_2016}. Out of $\tau_c$ samples, $\tau_u$ samples are used for the uplink. We assume perfect channel state information (CSI) at the APs, which is possible with a unique pilot per user and high enough transmission power during pilot transmission.
A compressed version of the received vector $\hat{\mathbf{y}}_l$ is stored in the local memory of AP $l$ and is defined as follows:
\begin{equation}
\begin{aligned}
\hat{\mathbf{y}}_l=\mathbf{y}_l+\mathbf{q}_l=\mathbf{H}_l\mathbf{s}+\mathbf{z}_l,
\end{aligned}
\label{eq2}
\end{equation}
where $\mathbf{z}_l=\mathbf{n}_l+\mathbf{q}_l$ is a spatially correlated noise vector with zero mean and covariance matrix $\mathbf{Z}_l$.
 The network-wide compressed received signal and noise vector can be expressed as $\hat{\mathbf{y}}=\begin{bmatrix}
    \hat{\mathbf{y}}_1^{\text{T}}& \hdots &\hat{\mathbf{y}}_L^{\text{T}}
\end{bmatrix}^{\text{T}}$ and $\mathbf{z}=\begin{bmatrix}
    \mathbf{z}_1^{\text{T}}& \hdots &\mathbf{z}_L^{\text{T}}
\end{bmatrix}^{\text{T}}$, respectively. The noise vectors in different APs are assumed to be independent, i.e., $\mathbf{Z}=\text{blkdiag}(
    \mathbf{Z}_1,\hdots,\mathbf{Z}_L)
$.

Algorithm \ref{algSRLS} summarizes the RLS steps for sequential uplink signal estimation among APs \cite{ranjbar2022}. Note that 
the superscript $n$ in algorithm \ref{algSRLS} differentiates the uplink samples in one coherence block. However, this superscript is not used anywhere else in the paper for notational simplicity.
\begin{algorithm}[h!]
\caption{ RLS algorithm for users' signal estimation}
 \label{algSRLS}
\begin{algorithmic}[1]
 \STATE \textbf{Initialize:}
 \STATE \hspace*{\algorithmicindent}\parbox[t]{.8\linewidth}{\raggedright  $\mathbf{\Gamma}_0=p\mathbf{I}_K$}
 \STATE \hspace*{\algorithmicindent}\parbox[t]{.8\linewidth}{\raggedright   $\hat{\mathbf{s}}^n_{0}=\mathbf{0}_{K\times 1}, \forall n\in[1:\tau_u]$}
\FOR{$l = 1 \dots L$}
\STATE{$\mathbf{\Gamma}_l=\mathbf{\Gamma}_{l-1}-\mathbf{\Gamma}_{l-1}\mathbf{H}_l^{\text{H}}\mathbf{Z}_l^{-\text{H}/2}(\mathbf{I}_N+\mathbf{Z}_l^{-1/2}\mathbf{H}_l\mathbf{\Gamma}_{l-1}\mathbf{H}_l^{\text{H}}\mathbf{Z}_l^{-\text{H}/2})^{-1}\mathbf{Z}_l^{-1/2}\mathbf{H}_l\mathbf{\Gamma}_{l-1}^{\text{H}}$}
\FOR{$n = 1 \dots \tau_u$} 
        \STATE{$
    \hat{\mathbf{s}}^n_l=\hat{\mathbf{s}}^n_{l-1}+\mathbf{\Gamma}_l\mathbf{H}_l^{\text{H}}\mathbf{Z}_l^{-\text{H}/2}(\hat{\mathbf{y}}^n_l-\mathbf{Z}_l^{-1/2}\mathbf{H}_l\hat{\mathbf{s}}^n_{l-1}).
    $}
               
\ENDFOR
\ENDFOR
\end{algorithmic}
\end{algorithm}
By updating the estimates of the users' signal as in algorithm \ref{algSRLS}, the users' signal estimates at the final AP will be:
 \begin{equation}
     \hat{\mathbf{s}}=(\mathbf{H}^{\text{H}}\mathbf{Z}^{-1}\mathbf{H}+\frac{1}{p}\mathbf{I}_k)^{-1}\mathbf{H}^{\text{H}}\mathbf{Z}^{-1}\hat{\mathbf{y}}.
 \label{eq3}
  \end{equation}

In a sequential fronthaul, the number of received signal vectors to be stored grows linearly with the number of APs in the network and should be processed at every symbol time. Thus, this memory should be very fast so that every sample can be processed promptly. On the contrary, the local CSI should be stored only once per coherence block in each AP.
Section \ref{sec2B} elaborates on storing the local received signal vectors in the limited memory. 
In Section \ref{sec3}, the compression of the local received signal vector at each AP is defined from a sum-SE optimization problem constrained by the limited memory capacity at the AP. 

\subsection{Storage of received signal vectors in the limited memory APs}\label{sec2B}
\begin{figure}[t]
    \centering
    \includegraphics[scale=0.5]{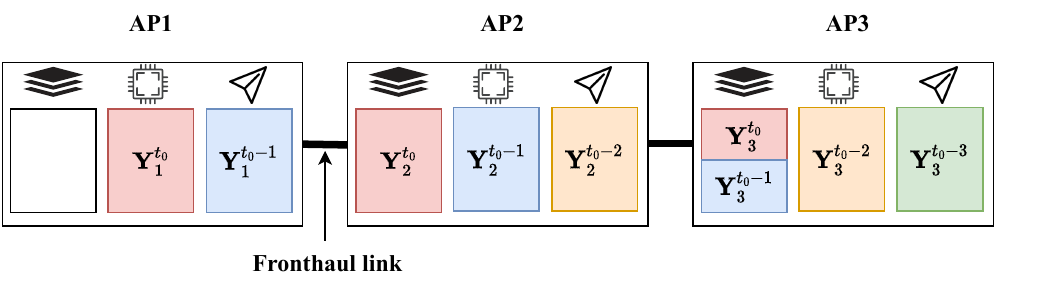}
    \caption{Sequential processing and storage in a daisy chain fronthaul topology }
    \label{fig2_seq}
\end{figure}

In a daisy chain fronthaul, the APs estimate each user's signal based on their local received signal vector and the signal estimates they receive from the previous AP, as shown in line (7) of algorithm \ref{algSRLS}. Therefore, they need to store their local received signal vector until the previous AP has finished processing its corresponding local received signal vector.
To make the problem more tangible, consider an orthogonal frequency-division
multiplexing (OFDM) system where the bandwidth is divided between $F$ subcarriers. Each AP receives $N$ OFDM symbols, one for each antenna. 
Suppose that $\mathbf{Y}_l^{t_0}\in\mathbb{C}^{N\times F}$ is a symbol matrix with each column corresponding to the received signal vector at AP $l$ for a particular subcarrier of symbol $t_0$, as shown in Fig~\ref{fig2_seq}.
When AP $1$ is processing $\mathbf{Y}_1^{t_0}$, the rest of the APs have their corresponding matrices, i.e., $\mathbf{Y}_l^{t_0}, \forall l\in\{2,\hdots, L\}$ in their memory. Similarly, when AP $2$ is processing $\mathbf{Y}_2^{t_0-1}$, the corresponding matrix at AP $l$, 
 i.e., $\mathbf{Y}_l^{t_0-1}, \forall l\in\{3,\hdots,L\}$, is stored in the memory. Accordingly, the number of the symbol matrix stored at AP $l$ is $l-1$ meaning that the number of received signal vectors stored at AP $l$ is $(l-1)F$, which increases linearly from one AP to the next by $F$. It is worth mentioning that processors are designed to process at least one symbol during a symbol duration to have a stable system. In other words, the rate of the locally received signal vectors entering the memory should be lower or, in the worst case, the same as the rate at which the processor is processing them. Therefore, the number of vectors stored in the AP's memory increases by $F$ from one AP to the next. 

 The memory to store the vectors is usually a fast on-chip cache memory \cite{jesusthesis}, close to the processing unit \cite{arm}.

\section{Received signal vector compression}\label{sec3}
In this section, for storing the received signal vectors, we consider 1) Joint compression of the received signal vector elements, also called vector-wise compression (VC) of the received signal vector, and 2) Element-wise compression (EC) of the received signal vector.
\subsection{Option 1: Vector-wise compression (VC) of the received signal vector}\label{sec3_A}
The compressed vector at AP $l$ can be represented as $\hat{\mathbf{y}}_{vl}$ and the relation between $\mathbf{y}_l$ and $\hat{\mathbf{y}}_{vl}$ follows as \cite{CovTho,jointKang}:
\begin{equation}
\hat{\mathbf{y}}_{vl}=\mathbf{y}_l+\mathbf{q}_{vl}=\mathbf{H}_l\mathbf{s}+\mathbf{n}_l+\mathbf{q}_{vl},
  \label{eq4}
\end{equation}
where $\mathbf{q}_{vl} \sim \mathcal{CN}(\mathbf{0}, \mathbf{Q}_{vl})$ represents the compression noise which is independent of $\mathbf{y}_l$. We also define $\mathbf{z}_{vl}=\mathbf{n}_l+\mathbf{q}_{vl}$, with covariance matrix $\mathbf{Z}_{vl}=\mathbb{E}\{\mathbf{z}_{vl}\mathbf{z}_{vl}^{\text{H}}\}=\mathbf{Q}_{vl}+\sigma^2\mathbf{I}_N$.
The relation between the number of bits to compress the received signal vector, $C_{s}$, and the compression noise covariance matrix $\mathbf{Q}_{vl}$ at AP $l$, conditioned on the local CSI is as follows:
\begin{equation}
\begin{aligned}
    C_{s}& =I(\mathbf{y}_l;\hat{\mathbf{y}}_{vl}|\mathbf{H}_l) \\
    & =\mathcal{H}(\hat{\mathbf{y}}_{vl}|\mathbf{H}_l)-\mathcal{H}(\hat{\mathbf{y}}_{vl}|\mathbf{y}_l,\mathbf{H}_l) \\
    & = \log_2 \det({\mathbf{Q}_{vl}^{-1}(p\mathbf{H}_l\mathbf{H}_l^{\text{H}}+\sigma^2\mathbf{I}_N)+\mathbf{I}_N}).
    \label{eq5}
\end{aligned}
\end{equation}
The network-wide compressed received vector is given as:
\begin{equation}
    \hat{\mathbf{y}}_v=\mathbf{y}+\mathbf{q}_v=\mathbf{H}\mathbf{s}+\underbrace{\mathbf{n}+\mathbf{q}_v}_{\mathbf{z}_v},
\label{eq6}
\end{equation}
where $\hat{\mathbf{y}}_v=\begin{bmatrix}
        \hat{\mathbf{y}}^{\text{T}}_{v1}& \hdots& \hat{\mathbf{y}}^{\text{T}}_{vL}
    \end{bmatrix}^{\text{T}}$, $\mathbf{q}_v=\begin{bmatrix}
        \mathbf{q}^{\text{T}}_{v1}& \hdots & \mathbf{q}^{\text{T}}_{vL}
\end{bmatrix}^{\text{T}}$,
and $\mathbf{z}_v=\begin{bmatrix}\mathbf{z}_{v1}^{\text{T}}&\hdots&\mathbf{z}_{vL}^{\text{T}}\end{bmatrix}^{\text{T}}$ is the receiver plus compression noise vector with covariance matrix $\mathbf{Z}_v=\mathbb{E}\{\mathbf{z}_{v}\mathbf{z}_{v}^{\text{H}}\}=\text{blkdiag}(\mathbf{Z}_{v1},\hdots,\mathbf{Z}_{vL})=\text{blkdiag}(\mathbf{Q}_{v1}+\sigma^2\mathbf{I}_N,\hdots,\mathbf{Q}_{vL}+\sigma^2\mathbf{I}_N)$. Following the discussion in Section \ref{secII-A}, the LS estimates of the users' signal can be formulated as below:
\begin{equation}
 \hat{\mathbf{s}}_v= \mathbf{V}_v\hat{\mathbf{y}}_v, 
 \label{eq7}
\end{equation}
where combiner matrix $\mathbf{V}_v$ is given as follows:
\begin{equation}
\mathbf{V}_v=(\mathbf{H}^{\text{H}}\mathbf{Z}_v^{-1}\mathbf{H}+\frac{1}{p}\mathbf{I}_k)^{-1}\mathbf{H}^{\text{H}}\mathbf{Z}_v^{-1}.
\label{eq8}
\end{equation} 
Having the estimates of 
the users' signal as in (\ref{eq7}),  the sum-SE of users is formulated as follows:
\begin{equation}
\begin{aligned}
    R_{VC}=&\frac{\tau_u}{\tau_c}\mathcal{I}(\mathbf{V}_v\hat{\mathbf{y}};\mathbf{s}|\mathbf{V}_v,\mathbf{H})\\=&\frac{\tau_u}{\tau_c}\Bigl(\mathcal{H}(\mathbf{V}_v{\hat{\mathbf{y}}}_v|\mathbf{V}_v,\mathbf{H})-\mathcal{H}(\mathbf{V}_v{\hat{\mathbf{y}}}_v|\mathbf{s},\mathbf{V}_v,\mathbf{H})\Bigr)\\=&\frac{\tau_u}{\tau_c}\log_2\det(p\mathbf{H}\mathbf{H}^{\text{H}}\mathbf{Z}_v^{-1}+\mathbf{I}_{NL})\\
    \myineqla&\frac{\tau_u}{\tau_c}\log_2 \prod_{l=1}^{L}\det(p\mathbf{H}_l\mathbf{H}_l^{\text{H}}\mathbf{Z}_{vl}^{-1}+\mathbf{I}_{N})\\
    =&\frac{\tau_u}{\tau_c}\sum_{l=1}^{L} \log_2\det(p\mathbf{H}_l\mathbf{H}_l^{\text{H}}\mathbf{Z}_{vl}^{-1}+\mathbf{I}_{N})\\=&\frac{\tau_u}{\tau_c}
    \sum_{l=1}^{L} \log_2\det(p\mathbf{H}_l\mathbf{H}_l^{\text{H}}(\mathbf{Q}_{vl}+\sigma^2\mathbf{I}_N)^{-1}+\mathbf{I}_{N}),
   \end{aligned}
    \label{eq9}  
\end{equation}
where the inequality $\myineqla$ holds due to the fact that matrix $\mathbf{Z}_v$
is a block-diagonal matrix, and the fact that the determinant of the positive (semi-) definite matrix is always smaller than the determinant of a diagonal matrix with the same diagonal elements \cite{horn_johnson_2012}. The detailed proof is omitted due to space limitations. The upper bound in equation (\ref{eq9}) is for the instantaneous sum-SE in a particular coherence block. 


A maximization problem can be formulated to find the optimal $\mathbf{Q}_{vl}, \forall l$ that maximizes the upper bound defined in (\ref{eq9}).

The upper bound on the sum-SE in (\ref{eq9}) is  
a summation of $L$ functions  
each of which depends only on the compression plus receiver noise covariance matrix of a single AP. Additionally,  each AP compresses its received signal vector in isolation from other APs. Therefore, maximization of the upper bound 
can be decomposed into $L$ smaller optimization problems to be solved at $L$ APs. Accordingly, the 
sum-SE maximization problem at AP $l$ is defined as follows:
\begin{equation}
     \begin{aligned}
\arg \max_{\mathbf{Q}_{vl}^{-1}\succeq 0} \quad & \log_2
    \det(p\mathbf{H}_l\mathbf{H}_l^{\text{H}}(\mathbf{Q}_{vl}+\sigma^2\mathbf{I}_N)^{-1}+\mathbf{I}_N)\\
\textrm{s.t.} \quad & C_{s}=\log_2 \det({\mathbf{Q}_{vl}^{-1}(p\mathbf{H}_l\mathbf{H}_l^{\text{H}}+\sigma^2\mathbf{I}_N)+\mathbf{I}_N}).
\end{aligned} 
\label{p10}
    \end{equation}
Similar to \cite{jointKang}  and \cite[app. B]{dis_comp}, the problem in (\ref{p10}) can be converted to an equivalent optimization problem that maximizes the objective function with respect to the eigenvalues of $\mathbf{Q}_{vl}^{-1}$. Due to space limitations, the proof is omitted. 

The optimal matrix $\mathbf{Q}_{vl}^{-1}$ is found to be as follows:
\begin{equation}
    \mathbf{Q}_{vl}^{-1*}= \mathbf{U}_l\mathbf{\Sigma}_{vlq}^{-1*}\mathbf{U}_l^{\text{H}},
    \label{eq11}
\end{equation}
where the columns of $\mathbf{U}_l\in \mathbb{C}^{N\times N}$ are the eigen vectors of $\mathbf{H}_l\mathbf{H}_l^{\text{H}}$ and the $i^{th}$ diagonal element of $\mathbf{\Sigma}_{vlq}^{-1*}$ is found as follows:

\begin{equation}
\begin{aligned}
       \lambda^*_{vlqi}=[\frac{1}{\mu^*}(\frac{1}{\sigma^2}-\frac{1}{p\lambda_{li}^2+\sigma^2})-\frac{1}{\sigma^2}]^+, \forall i,
        \label{eq12}
        \end{aligned}
    \end{equation}
    where $\lambda_{li}^2, \forall i\in\{1,\hdots, N\}$ are the eigenvalues of  $\mathbf{H}_l\mathbf{H}^H_l$. Having $\mathbf{Q}_{vl}^{-1*}$, $\mathbf{Q}_{vl}^{*}$ can be determined accordingly. $\mu^*$ is the Lagrange multiplier and is found to satisfy the equality constraint in (\ref{p10}). 

\subsection{Option 2: Element-wise compression (EC) of the received signal vector}\label{sec3_B}
In the EC of the received vector, each element is compressed individually. 
The bits allocated to the compression of the $i^{th}$ element of the local received vector $\mathbf{y}_l$ at AP $l$ can be denoted by $b_{li}$ and $C_{s}=\sum_{i=1}^{N}b_{li}$. The compressed $i^{th}$ element is given as follows:
\begin{equation}
\hat{y}_{eli}=y_{li}+q_{eli}=\mathbf{H}_{l[i,:]}\mathbf{s}+n_{li}+q_{eli},
   \label{eq13}
\end{equation}
where the subscript $[i,:]$ 
specifies the $i^{th}$ row of matrix $\mathbf{H}_l$, $q_{eli}\sim \mathcal{C}\mathcal{N}(0,\sigma^2_{eli})$, $\mathbf{q}_{el}=\begin{bmatrix}
    {q}_{el1}&\cdots & {q}_{elN}
\end{bmatrix}^{\text{T}}$
is the compression noise vector with covariance matrix $\mathbf{Q}_{el}=\mathbb{E}\{\mathbf{q}_{el}\mathbf{q}_{el}^H\}$. We define $z_{eli}=n_{li}+q_{eli}$, $\mathbf{z}_{el}=\begin{bmatrix}
    {z}_{el1}&\cdots & {z}_{elN}
\end{bmatrix}^{\text{T}}$ with covariance matrix $\mathbf{Z}_{el}=\mathbb{E}\{\mathbf{z}_{el}\mathbf{z}_{el}^H\}=\mathbf{Q}_{el}+\sigma^2\mathbf{I}_N$. Furthermore, $n_{li}$ is the $i^{th}$ element of the noise vector $\mathbf{n}_l$ and $\hat{\mathbf{y}}_{el}=\begin{bmatrix}
    \hat{y}_{el1}& \hdots&\hat{y}_{elN}
    \end{bmatrix}^{\text{T}}$ is the compressed received signal vector.
The relation between $b_{li}$ and compression noise of the $i^{th}$ element is:
\begin{equation}
\begin{aligned}
    b_{li}&=\mathcal{I}(y_{eli};\hat{y}_{eli}|\mathbf{H}_{l[i,:]})\\&=\mathcal{H}(\hat{y}_{eli}|\mathbf{H}_{l[i,:]})-\mathcal{H}(\hat{y}_{eli}|y_{eli},\mathbf{H}_{l[i,:]})\\&=\log_2(\frac{p\lVert \mathbf{H}_{l[i,:]}\rVert^2+\sigma^2}{\sigma_{eli}^{2}}+1),
    \label{eq14}
    \end{aligned}
\end{equation}
The diagonal elements of the covariance matrix $\mathbf{Q}_{el}$ can be calculated based on value of $b_{li}, \forall i\in\{1,\hdots,N\}$ and represented as $\sigma_{eli}^{2}, \forall i\in\{1,\hdots,N\}$. 
The off-diagonal elements of matrix $\mathbf{Q}_{el}$ are unknown.

We define diagonal matrix $\mathbf{P}_l$ with the variance of the elements of the local received signal vector $\{\mathbf{y}_l|\mathbf{H}_l\}$ as its diagonal elements,
\begin{equation}
\begin{aligned}
    \mathbf{P}_l &=\text{diag}(\mathbb{E}\{\mathbf{y}_l\mathbf{y}_l^H|\mathbf{H}_l\})\\&= \text{diag}\left(p\lVert \mathbf{H}_{l[1,:]}\rVert^2+\sigma^2, \cdots, p\lVert \mathbf{H}_{l[N,:]}\rVert^2+\sigma^2\right),
    \label{eq15}
    \end{aligned}
\end{equation}
and the diagonal matrix $\mathbf{Q}_{el}^d$ with variance of the elements of the compression noise vector $\mathbf{q}_{el}$ as its diagonal elements, 
\textcolor{black}{\begin{equation}
    \mathbf{Q}_{el}^d = \text{diag}\left(\sigma_{el1}^{2}, \dots, \sigma_{elN}^{2}\right).
    \label{eq16}
\end{equation}}
We can relate $C_{s}$ to the variance of the compression noise vector elements as below:
\begin{equation}
\begin{aligned}
    C_{s} & =\sum_{i=1}^{N}b_{li}=\log_2\prod_{i=1}^N(\frac{p\lVert \mathbf{H}_{l[i,:]}\rVert^2+\sigma^2}{\textcolor{black}{\sigma_{eli}^{2}}}+1) \\
    & =\log_2\det( {(\mathbf{Q}^{d}_{el})}^{-1}\mathbf{P}_l+\mathbf{I}_N). 
    \label{eq17}
\end{aligned}
\end{equation}

The network-wide compressed received vector is as 
 follows:
 \begin{equation}
\hat{\mathbf{y}}_{e}=\mathbf{y}+\mathbf{q}_e=\mathbf{H}\mathbf{s}+\underbrace{\mathbf{n}+\mathbf{q}_e}_{\mathbf{z}_e},
\label{eq18}
\end{equation}
where $\hat{\mathbf{y}}_{e}=\begin{bmatrix}
        \hat{\mathbf{y}}^{\text{T}}_{e1}& \hdots& \hat{\mathbf{y}}_{eL}^{\text{T}}
    \end{bmatrix}^{\text{T}}$, $\mathbf{q}_e=\begin{bmatrix}
        \mathbf{q}_{e1}^{\text{T}}& \hdots& \mathbf{q}_{eL}^{\text{T}}
\end{bmatrix}^{\text{T}}$ and $\mathbf{z}_e=\begin{bmatrix}\mathbf{z}_{e1}^{\text{T}}&\hdots&\mathbf{z}_{eL}^{\text{T}}\end{bmatrix}^{\text{T}}$ with covariance matrix $\mathbf{Z}_e=\text{blkdiag}(\mathbf{Z}_{e1},\hdots,\mathbf{Z}_{eL})=\text{blkdiag}(\mathbf{Q}_{e1}+\sigma^2\mathbf{I}_N,\hdots,\mathbf{Q}_{eL}+\sigma^2\mathbf{I}_N)$.
In EC, the correlation between compression noise elements at one AP, i.e., the off-diagonal elements of $\mathbf{Q}_{el}, \forall l$, are unknown.
 Therefore, while computing the combining vector to estimate 
 users' signal, the correlation of compression noise elements at each AP is ignored ($\mathbf{Q}_{el}, \forall l$ is assumed to be a diagonal matrix), which will adversely affect the estimation quality.
The sequential estimation of 
user signals results in equations (\ref{eq19}) and (\ref{eq20}) with $\mathbf{Z}_v$ in (\ref{eq8}) replaced with the diagonal matrix $\mathbf{Z}_e^d=\text{blkdiag}(\mathbf{Z}_{e1}^d,\hdots,\mathbf{Z}_{eL}^d)=\text{blkdiag}(\mathbf{Q}_{e1}^d+\sigma^2\mathbf{I}_N,\hdots,\mathbf{Q}_{eL}^d+\sigma^2\mathbf{I}_N)$,
\begin{equation}
 \hat{\mathbf{s}}_e= \mathbf{V}_e\hat{\mathbf{y}}_e,
 \label{eq19}
\end{equation}
where combining matrix $\mathbf{V}_e$ is formulated as follows:
\begin{equation}
\mathbf{V}_e=(\mathbf{H}^{\text{H}}{(\mathbf{Z}_e^d)}^{-1}\mathbf{H}+\frac{1}{p}\mathbf{I}_K)^{-1}\mathbf{H}^{\text{H}}{(\mathbf{Z}_e^d)}^{-1}.
\label{eq20}
\end{equation}

 The maximization of the 
 sum-SE in option 2 follows the same steps as in Section \ref{sec3_A}. 
 The user's sum-SE using EC can
     be simplified as follows:
    \begin{equation}
\begin{aligned}
R_{EC}&=\frac{\tau_u}{\tau_c}\log_2\det(p\mathbf{H}\mathbf{H}^{\text{H}}(\mathbf{Z}^d_e)^{-1}+\mathbf{I}_{NL})\\&\myineqla\frac{\tau_u}{\tau_c}\sum_{l=1}^L\log_2\det(p\mathbf{H}_l\mathbf{H}_l^{\text{H}}(\mathbf{Z}^d_{el})^{-1}+\mathbf{I}_{N})\\&\myineqlb\frac{\tau_u}{\tau_c}\sum_{l=1}^L\log_2\det(p\text{diag}(\mathbf{H}_l\mathbf{H}_l^{\text{H}})(\mathbf{Z}^d_{el})^{-1}+\mathbf{I}_{N})\\&=\frac{\tau_u}{\tau_c}\sum_{l=1}^L\log_2\det(p\mathbf{W}_l(\mathbf{Q}_{el}^d+\sigma^2\mathbf{I})^{-1}+\mathbf{I}_{N}),
   \end{aligned}
    \label{eq21}  
\end{equation}
where $\mathbf{W}_l$ is defined as
    $\mathbf{W}_l=\text{diag}(\lVert\mathbf{H}_{l[1,:]}\rVert^2,\hdots,{\lVert\mathbf{H}_{l[N,:]}}\rVert^2)
        $.
    In (\ref{eq21}), $\myineqla$ and $\myineqlb$ are proved similar to $\myineqla$ in (\ref{eq9}). 
    
    The optimization problem to find the diagonal elements of $\mathbf{Q}_{el}^d$ and, subsequently, the number of bits to compress each of the elements of the local received signal vector $\mathbf{y}_l$ is formulated as follows:
    \begin{equation}
     \begin{aligned}
\arg \max_{({\mathbf{Q}^d_{el})}^{-1}\succeq 0} \quad &  \log_2
    \det(p\mathbf{W}_l(\mathbf{Q}_{el}^d+\sigma^2\mathbf{I}_N)^{-1}+\mathbf{I}_N)\\
\textrm{s.t.} \quad & C_{s}=\log_2 \det({({\mathbf{Q}_{el}^d})}^{-1}\mathbf{P}_l+\mathbf{I}_N),
\end{aligned} 
\label{p23}
\end{equation}
with $\mathbf{P}_l$ defined in (\ref{eq15}). Note that $\mathbf{P}_l=p\mathbf{W}_l+\sigma^2\mathbf{I}_N$.
Following a similar derivation as in Section \ref{sec3_A}, the $i^{th}$ diagonal element of matrix $({{\mathbf{Q}_{el}^d}})^{-1}$, shown as $\frac{1}{\sigma^2_{eli}}$,  can be calculated as follows:
\begin{equation}
    \frac{1}{\sigma^2_{eli}}=[\frac{1}{\mu^*}(\frac{1}{\sigma^2}-\frac{1}{{\mathbf{P}_l}_{[i,i]}})-\frac{1}{\sigma^2}]^+.
    \label{eq24}
\end{equation}

\section{Memory capacity model}\label{sec4}
Regarding the memory capacity at the APs, two general scenarios are considered.
\begin{itemize}
    \item \textbf{Fixed per AP (FAP)}: There is a fixed memory capacity $C_{AP}$ per AP. Therefore, the total memory capacity depends on the number of APs.
    \item \textbf{Fixed total (FT)}: There is a fixed total memory capacity $C_{T}$ that is divided among APs. Therefore, the memory capacity allocated to each AP depends on the number of APs.
\end{itemize}
We assume that received signal vectors are stored in on-chip cache memory\cite{jesusthesis}. Cache memory is desirable for its fast accessibility and low energy consumption. In reality, local CSI should also be stored in the memory. However, as the amount of data related to CSI is similar in each AP, we ignore the low precision storage of the local CSI. 
 Note that the memory capacity available at one AP is shared among all the sub-carriers.

\section{simulation results}
This section presents simulation results, which give insight into how the limited memory capacity in each AP can affect the average per-user SE. The simulation area is square with a perimeter of $D=500\text{m}$. The APs are located on the perimeter of the area, and the distance between any two APs is the same. The users are uniformly located in a concentric square with a perimeter of $400\text{m}$. The vertical distance between a user and an AP is $5\text{m}$ \cite{Shaik2021}.
The total number of antennas is $\textcolor{black}{NL=128}$ which are distributed in $\textcolor{black}{L=\{2,4,8,16,32,64,128\}}$ APs. 
The path-loss model of an urban microcell with 2GHz carrier frequency is considered \cite{3gpp_PL,Shaik2021}. Accordingly, the large-scale fading coefficient is defined as follows:
\begin{equation}
    \beta_{kl}=-30.5-36.7\log_{10}(\frac{d_{kl}}{1\text{m}}),
\end{equation}
where $d_{kl}$ and $\beta_{kl}$ are the distance and large scale fading coefficient between user $k$ and AP $l$, repectively. The noise variance at the APs is $\sigma^2=-85\text{dBm}$, and the users' transmit power is $p=10\text{mWatt}$.
\begin{figure}[!h]
    \centering
    \pgfplotsset{width=6.7cm,compat=1.18}
\pgfplotsset{every x tick label/.append style={font=\small, yshift=0.5ex},every y tick label/.append style={font=\small, xshift=0.5ex},
every axis legend/.append style={
at={(1.05,2)},
anchor=northwest,font=\small
}}

\begin{tikzpicture}[scale=0.7]
\begin{scope}[xshift=0cm,yshift=0cm]
\begin{axis}[
domain=0:4,
grid=major,
ylabel= Average per-user SE (bit/sec/Hz),
xlabel= Number of the APs,
xmin=1,
xmax=7,
xtick style={color=clr3},
 xtick={1,2,3,4,5,6,7},
 xticklabels={L=2,L=4,L=8,L=16,L=32,L=64,L=128},
ymin=2,
ymax=8,
ytick={3,4,5,6,7,8},
mark size=4.0 pt,
legend columns=2,
legend style={nodes={scale=0.9, transform shape},at={(1.05,-0.55)},anchor=south},
]
        
        

        \addplot [cyan, mark=o]
        table[x expr=\coordindex+1,y=B,col sep=comma] {Data/Data_seventh_run_fixed_PER_AP/data_K4_cfinf.csv};
        \addlegendentry{Infinite memory capacity}

        \addplot [black, mark=star]
        table[x expr=\coordindex+1,y=B,col sep=comma] {Data/DATA_UPLOADED_form_journal/data_K4_cf256KB.csv};
        \addlegendentry{FAP, VC, $C_{AP}=256KB$}

        \addplot [black, mark=triangle]
        table[x expr=\coordindex+1,y=C,col sep=comma] {Data/DATA_UPLOADED_form_journal/data_K4_cf256KB.csv};
        \addlegendentry{FAP, EC, $C_{AP}=256KB$}
        
         \addplot [red, mark=star]
        table[x expr=\coordindex+1,y=B,col sep=comma] {Data/DATA_UPLOADED_form_journal/data_K4_cf64KB.csv};
        \addlegendentry{FAP, VC, $C_{AP}=64KB$}

        \addplot [red, mark=triangle]
        table[x expr=\coordindex+1,y=C,col sep=comma] {Data/DATA_UPLOADED_form_journal/data_K4_cf64KB.csv};
        \addlegendentry{FAP, EC, $C_{AP}=64KB$}
\end{axis}
\end{scope}
 \begin{scope}[xshift=5.9cm,yshift=0cm]
\begin{axis}[
domain=0:4,
grid=major,
xlabel= Number of the APs,
xmin=1,
xmax=7,
xtick style={color=clr3},
 xtick={1,2,3,4,5,6,7},
 xticklabels={L=2,L=4,L=8,L=16,L=32,L=64,L=128},
ymin=2,
ymax=8,
ytick={3,4,5,6,7,8},
mark size=4.0 pt,
legend columns=1,
legend style={nodes={scale=0.9, transform shape},at={(0.5,-0.39)},anchor=south},
]
        \addplot [cyan, mark=o]
        table[x expr=\coordindex+1,y=B,col sep=comma] {Data/Data_fifthrun_fixed_PER_AP/data_K64_cfinf.csv};

        \addplot [black, mark=star]
        table[x expr=\coordindex+1,y=B,col sep=comma] {Data/DATA_UPLOADED_form_journal/data_K64_cf256KB.csv};

       \addplot [black, mark=triangle]
        table[x expr=\coordindex+1,y=C,col sep=comma] {Data/DATA_UPLOADED_form_journal/data_K64_cf256KB.csv};
        
         \addplot [red, mark=star]
        table[x expr=\coordindex+1,y=B,col sep=comma] {Data/DATA_UPLOADED_form_journal/data_K64_cf64KB.csv};

        \addplot [red, mark=triangle]
        table[x expr=\coordindex+1,y=C,col sep=comma] {Data/DATA_UPLOADED_form_journal/data_K64_cf64KB.csv};


\end{axis}
\end{scope}

\end{tikzpicture}
    \caption{Average per-user SE comparison using FAP memory model in a daisy chain fronthaul topology with two different numbers of users: (Left) $K=4$. (Right) $K=64$.}
    \label{figlimited_vs_inf}
\end{figure}
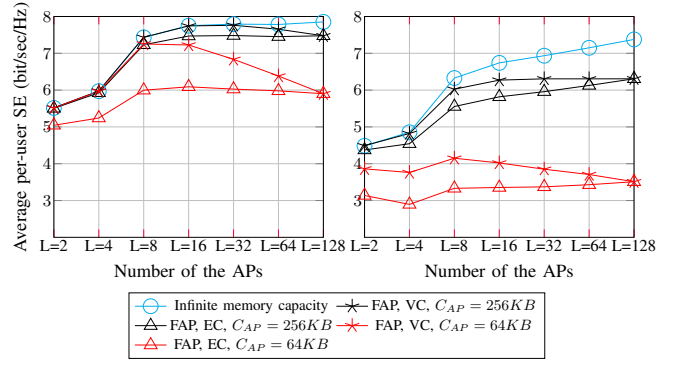


In this section, $\frac{\mathbb{E}\{R_{VC}\}}{K}$ and $\frac{\mathbb{E}\{R_{EC}\}}{K}$ versus the number of APs are plotted. The expectations are with respect to all kinds of randomness. Scaling factor $\frac{\tau_u}{\tau_c}$ exists in both (\ref{eq9}) and (\ref{eq21}), and as we don't consider any specific values for $\tau_u$ and $\tau_c$, the factor is omitted while plotting the simulation results.

To make the simulation results clear, we start with an example. In the simulation figures, in case of $\{L=64, N=2\}$, the last AP stores $63F$ \textbf{two}-dimensional received signal vectors in the memory, and in case of $\{L=128, N=1\}$, the last AP stores $127F$ received signal \textbf{scalars} (as there is one antenna per AP), according to Section \ref{sec2B}.
The total number of scalars to be stored in the last AP in the two cases mentioned are close to each other (i.e., $2\times63F$ versus $127F$ ). Therefore, it would seem that the compression noise should be almost the same. 
However, using VC and in the case of $\{L=64, N=2\}$, the last AP compresses scalars two by two ($63F$ pair of scalars), and in the case of $\{L=128, N=1\}$, the last AP compresses each scalar individually.
Jointly compressing every two scalars in $\{L=64, N=2\}$ allows the AP to use the available memory more efficiently than $\{L=128, N=1\}$. Consequently, when the memory capacity is limited (e.g., $C_{AP}=64KB$ in FAP), even though the number of scalars to be stored in the memory of the last AP in both cases is almost the same, $\{L=64, N=2\}$ outperforms $\{L=128, N=1\}$  even with the reduced macro-diversity. In other words, macro diversity in the case of $\{L=128, N=1\}$ can not compensate for the adverse effect of compression noise on average per-user SE. The above comparison can also be made between any other values of $L$. A similar conclusion can be drawn for EC. 

In Fig.~\ref{figlimited_vs_inf}, 
it is observed that,
 under the assumption of infinite memory capacity, the distribution of the antennas in single-antenna APs improves the average per-user SE, especially when the number of users is large, e.g., $K=64$. However, with the realistic assumption of limited memory capacity in each AP and using VC, the distribution of the antennas in single-antenna APs not only does not help in average per-user SE improvement but also reduces the average per-user SE, e.g., when $C_{AP}=64KB$.
\begin{figure}[h]
    \centering
    \pgfplotsset{width=6.7cm,compat=1.18}
\pgfplotsset{every x tick label/.append style={font=\small, yshift=0.5ex},every y tick label/.append style={font=\small, xshift=0.5ex},
every axis legend/.append style={
at={(1.02,1)},
anchor=north west,font=\small
}}

\begin{tikzpicture}[scale=0.7]
\begin{scope}[xshift=0cm,yshift=0cm]
\begin{axis}[
domain=0:4,
grid=major,
ylabel= Average per-user SE (bit/sec/Hz),
xlabel= Number of the APs,
xmin=1,
xmax=7,
xtick style={color=clr3},
 xtick={1,2,3,4,5,6,7},
 xticklabels={L=2,L=4,L=8,L=16,L=32,L=64,L=128},
ymin=1,
ymax=8,
ytick={2,3,4,5,6,7,8},
mark size=4.0 pt,
legend columns=2,
legend style={nodes={scale=0.9, transform shape},at={(1.05,-0.47)},anchor=south},
]
\addplot [black, mark=star]
        table[x expr=\coordindex+1,y=B,col sep=comma] {Data/DATA_UPLOADED_form_journal/data_K4_cf8MB.csv};
        \addlegendentry{FT, VC,  $C_{T}=8MB$}
        
         \addplot [black, mark=triangle]
        table[x expr=\coordindex+1,y=C,col sep=comma] {Data/DATA_UPLOADED_form_journal/data_K4_cf8MB.csv};
        \addlegendentry{FT, EC, $C_{T}=8MB$}
        \addplot [red, mark=star]
        table[x expr=\coordindex+1,y=B,col sep=comma] {Data/DATA_UPLOADED_form_journal/data_K4_cf32MB.csv};
        \addlegendentry{FT, VC, $C_{T}=32MB$}
        \addplot [red, mark=triangle]
        table[x expr=\coordindex+1,y=C,col sep=comma] {Data/DATA_UPLOADED_form_journal/data_K4_cf32MB.csv};
        \addlegendentry{FT, EC, $C_{T}=32MB$}

        \end{axis}
\end{scope}
        
       \begin{scope}[xshift=5.9cm,yshift=0cm]
\begin{axis}[
domain=0:4,
grid=major,
xlabel= Number of the APs,
xmin=1,
xmax=7,
xtick style={color=clr3},
 xtick={1,2,3,4,5,6,7},
 xticklabels={L=2,L=4,L=8,L=16,L=32,L=64,L=128},
ymin=1,
ymax=8,
ytick={2,3,4,5,6,7,8},
mark size=4.0 pt,
legend columns=1,
legend style={nodes={scale=0.9, transform shape},at={(1.6,-0.3)},anchor=north},
]

        \addplot [black, mark=star]
        table[x expr=\coordindex+1,y=B,col sep=comma] {Data/DATA_UPLOADED_form_journal/data_K64_cf8MB.csv};
        \addplot [black, mark=triangle]
        table[x expr=\coordindex+1,y=C,col sep=comma] {Data/DATA_UPLOADED_form_journal/data_K64_cf8MB.csv};
        \addplot [red, mark=star]
        table[x expr=\coordindex+1,y=B,col sep=comma] {Data/DATA_UPLOADED_form_journal/data_K64_cf32MB.csv};
        \addplot [red, mark=triangle]
        table[x expr=\coordindex+1,y=C,col sep=comma] {Data/DATA_UPLOADED_form_journal/data_K64_cf32MB.csv};

\end{axis}
\end{scope}

\end{tikzpicture}
    \caption{Average per-user SE Comparison using FT memory model in a daisy chain fronthaul topology with two different numbers of users:  
    (Left) $K=4$, (Right) $K=64$. 
    }
    \label{fig3}
\end{figure}
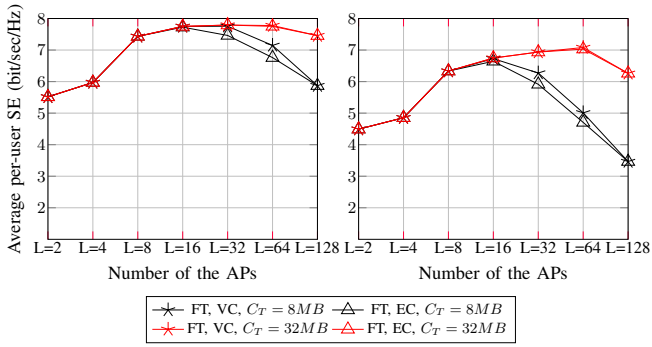
In Fig.~\ref{fig3}, FT memory model is considered, and a similar trend to Fig.~\ref{figlimited_vs_inf} is observed. Unlike FAP, in FT memory model, adding APs doesn't increase the total memory but makes the memory per AP smaller.
Furthermore, it is observed that the performance improvement of VC over EC is relatively small in this case, especially for the case of a low number of APs. This is because when the number of APs is low, each AP receives a large share of the total memory, reducing the difference between EC and VC.
\section{Conclusions}
This paper discusses 
sequential processing with limited memory APs in a cell-free massive MIMO network with daisy chain fronthaul topology. 
The paper shows a trade-off between achieving macro diversity by distributing the antennas as much as possible in more APs and reducing the adverse effect of compression noise by distributing the antennas in fewer APs. 
Specifically, based on simulation results, distributing the antennas can benefit the average per-user SE when there is no memory limit at the APs, especially when the number of users is high. However, this is not the case when we limit the memory available at each AP to store the received signal vectors. With limited memory capacity constraints at each AP, the antennas tend to be collocated in fewer APs. Hence, the memory capacity highly impacts the number of APs among which the available antennas should be distributed.
\section{Acknowledgement}
This work is supported by European Union’s
Horizon 2020 research and innovation program under grant
agreements: 101013425 (REINDEER) and 101017171 (MARSAL).

The resources and services used in this work were provided by the VSC (Flemish Supercomputer Center), funded by the Research Foundation - Flanders (FWO) and the Flemish Government.
\nocite{HardwareBjörnson}
\nocite{jointKang}
\nocite{masoumiperformance}
\nocite{Shaik2021}
\bibliographystyle{ieeetr}
\bibliography{refs}
\end{document}